\documentclass[manuscript=article]{achemso}
\usepackage{color}
\usepackage{soul}
\sethlcolor{yellow}
\usepackage{tabularx}
\usepackage{graphicx}
\usepackage{bm}
\usepackage{float}
\usepackage{lineno}
\usepackage[version=3]{mhchem}
\usepackage{soul}

\newcommand{\COMMENT}[1]{}
\newcommand{\textapprox}{{\raise.17ex\hbox{$\scriptstyle\mathtt{\sim}$}}}

\author{Hyeong-Seok D. Kim}
\affiliation{The Makineni Theoretical Laboratories, Department of Chemistry, University of Pennsylvania, Philadelphia, Pennsylvania 19104--6323, USA}
\author{Jing Yang}
\affiliation{The Makineni Theoretical Laboratories, Department of Chemistry, University of Pennsylvania, Philadelphia, Pennsylvania 19104--6323, USA}
\author{Yubo Qi}
\affiliation{The Makineni Theoretical Laboratories, Department of Chemistry, University of Pennsylvania, Philadelphia, Pennsylvania 19104--6323, USA}
\author{Andrew M. Rappe}
\affiliation{The Makineni Theoretical Laboratories, Department of Chemistry, University of Pennsylvania, Philadelphia, Pennsylvania 19104--6323, USA}
\date{\today}
\title{Adsorption of Benzene on the RuO$_2$(110) Surface}

\begin{document}


\newpage
\begin{tocentry}
\begin{figure}[H]
\begin{center}
\includegraphics[scale=0.17]{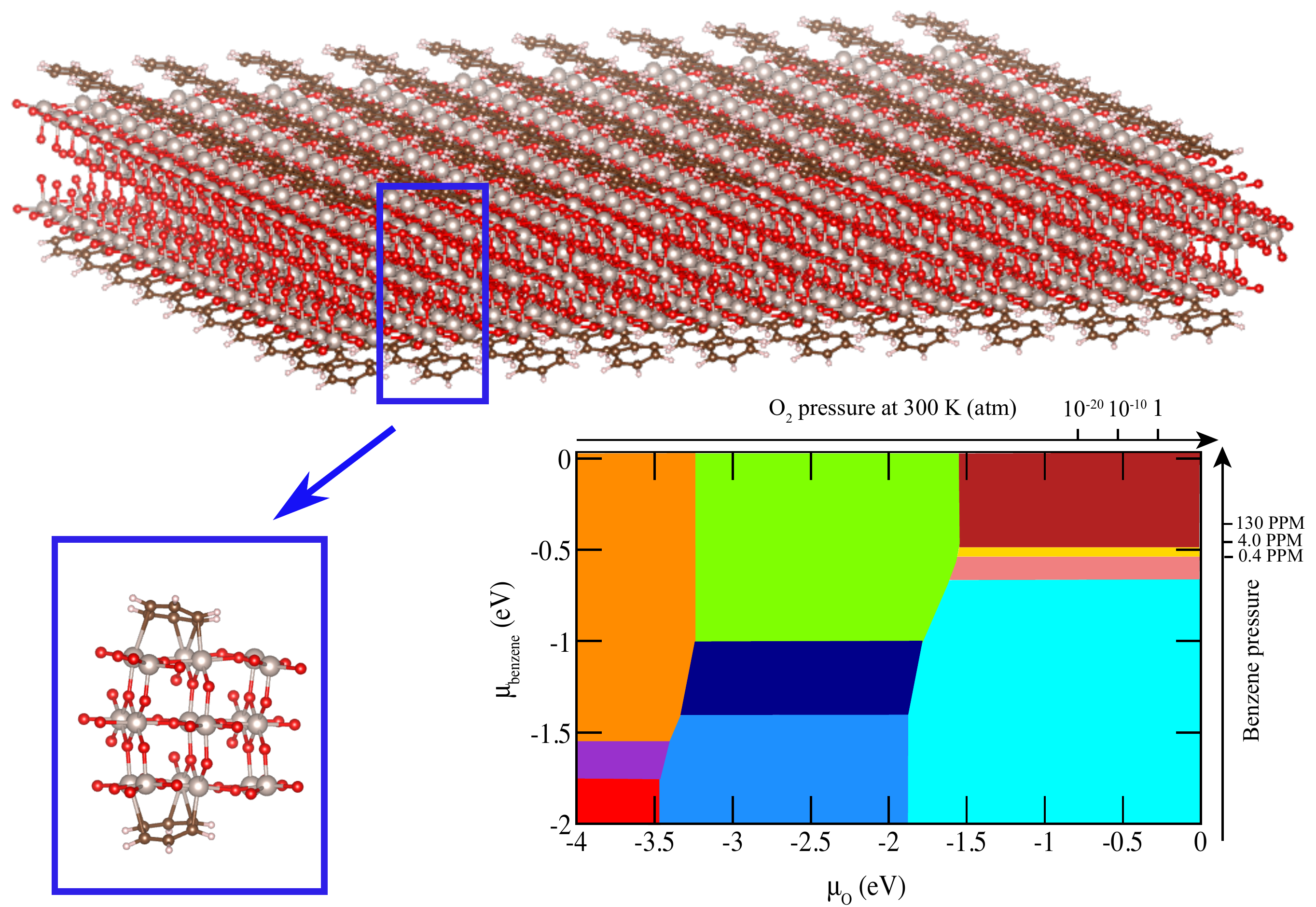}
\end{center}
\end{figure}
\end{tocentry}

\begin{abstract}  

Hydrocarbon tribopolymer, a type of polymer formed due to friction between surfaces, is a major impediment to the development of micro- and nanoelectromechanical systems (MEMS/NEMS) devices for industrial application. Tribopolymer buildup can prevent MEMS and NEMS from making or breaking electrical contact. We describe the adsorption of benzene (C$_6$H$_6$) on the RuO$_2$(110) surface using density functional theory. This adsorption is an important initial step in the mechanism of hydrocarbon tribopolymer layer formation on MEMS and NEMS devices. The adsorption interaction is studied by considering three oxygen coverages of RuO$_2$(110) and all the possible adsorption sites for benzene. We find that adsorption of benzene on O-poor RuO$_2$(110) via C-Ru bonds is stronger than adsorption on the O-rich RuO$_2$(110) via H-O bonds. For an in-depth study of the adsorption behavior, we include the van der Waals interaction for a holistic investigation. By incorporating the thermodynamic chemical potentials into the adsorption simulations, we describe a model that can provide guidance for realistic situations.

%
%


\end{abstract}

\section{Introduction}

Over the past twenty years, the RuO$_2$(110) surface has become a key model system for transition metal oxide catalysis\cite{Over02p37}. Its high electrical conductivity and high bulk modulus\cite{lundin98p4979} make this metal oxide a suitable material for application in microelectromechanical and nanoelectromechanical systems (MEMS/NEMS) devices. But as shown experimentally\cite{Brand13p1248}, hydrocarbon tribopolymer layers form as benzene is introduced into the atmosphere.
This tribopolymer increases the electrical contact resistance, which leads to electrical failure of the switches. This obstacle is yet to be solved, and the mechanism of tribopolymer formation remains unclear. 
Since aspects such as atmospheric composition and gas pressures have been shown to affect the contamination rate of the switches\cite{brand13p341}, studying the adsorption of gases on the contact surfaces will provide crucial understanding of the initial step of tribopolymer formation.
 
The surface catalytic properties of RuO$_2$(110) have been well studied for small molecules. Molecules including CO\cite{kim01p115419,reuter03p045407,reuter06p045433}, H$_2$O\cite{chu01p3364,lobo03p279}, O$_2$\cite{kim01p3752}, N$_2$\cite{kim01p115419}, methanol\cite{madhavaram01p296}, CO$_2$\cite{wang02p5476}, NO\cite{wang03p13918}, ethylene\cite{paulus04p989}, NH$_3$\cite{wang05p7883}, HCl\cite{over12p6779} and H$_2$\cite{knapp07p5363} are reported to be adsorbed from the gas phase directly to a single catalytically active atom on the surface. For benzene, there is a similar but slightly different mechanism of adsorption, in which the adsorbed molecule has a collective interaction with multiple surface atoms. 
This arises partly because the benzene molecule is a comparatively larger molecule than the previously studied ones, covering many surface atoms. This study of benzene chemisorption
is motivated by recent breakthroughs in NEMS devices and the
importance of discovering tribopolymer-resistant surfaces.  This
long-term goal requires understanding of the components and their
interactions. The complexity of benzene adsorption on RuO$_2$ due to the
large molecular size and the rich surface structure mandates in-depth
study and extends the recent literature on RuO$_2$ surface structure and
small-molecule chemisorption. 
 
In this work, we investigate the adsorption of benzene on the RuO$_2$(110) surface. Our proposed adsorption mechanisms incorporate the effect of varying benzene coverage and O content of the RuO$_2$ reconstruction. The surface terminations and oxygen coverages are determined by previous studies\cite{Reuter01p035406}, and adsorption Gibbs free energies were calculated by density functional theory (DFT) to study the thermodynamic properties of benzene adsorption under different conditions. In addition, we generate a phase diagram to predict the most favored benzene coverage.

%
%

\section{Computational Methods }

\subsection{DFT calculations}
In this investigation, density functional theory (DFT) calculations are performed using the Quantum ESPRESSO package \cite{Giannozzi09p395502} with designed nonlocal pseudopotentials \cite{rappe90p1227,ramer99p12471} from the OPIUM code \cite{Opium}. A 50 Ry plane wave cutoff is used, and the electronic exchange correlation energy is calculated with the generalized gradient approximation (GGA) of Perdew, Burke, and Ernzerhof \cite{Perdew96p3865}.
Later, the vdW correction methods, such as DFTD3~\cite{Grimme04p1463,Grimme06p1787,Grimme10p154104} and TS~\cite{Tkatchenko09p073005}, were applied to all the modeled systems.
An $8\times8\times8$ Monkhorst-Pack \cite{monkhorst76p5188} k-point mesh is used for variable cell relaxation of bulk RuO$_2$, and an $8\times8\times1$ k-point mesh is used for all the other calculations such as the slab, adsorption, and molecule relaxation studies.


\subsection{Surface structure and adsorption energy model systems}
\begin{figure}[ht]
\caption{Three possible terminations of the RuO$_2$ rutile (110) plane are studied: RuO$_2$(110)-Ru(left), stoichiometric RuO$_2$(110)-O$^{\text{bridge}}$(middle) and RuO$_2$(110)-O$^{\text{cus}}$(right).}
\begin{center}
\includegraphics[scale=0.20,angle=0]{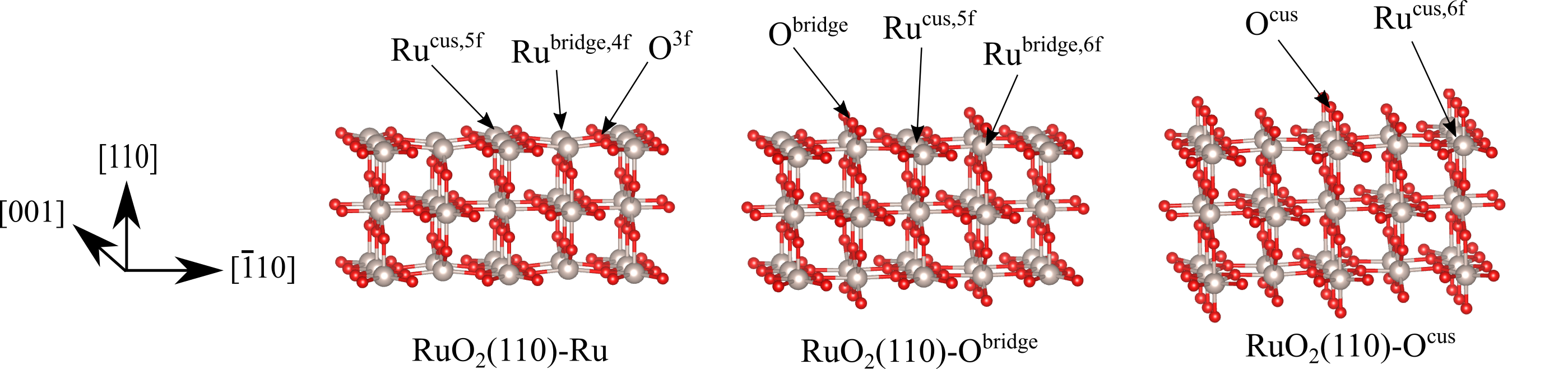}
\end{center}
\label{figure1}
\end{figure}

The stoichiometric RuO$_2$(110)-O$^{\text{bridge}}$ surface was previously considered to be the most stable surface at ambient conditions, but recent works show that the stable surface structure depends on the chemical potential of oxygen \cite{Reuter01p035406,reuter03p045407}. As shown in  FIG \ref{figure1}, three types of RuO$_2$(110) surface structures are stable: RuO$_2$(110)-Ru, RuO$_2$(110)-O$^{\text{bridge}}$, and RuO$_2$(110)-O$^{\text{cus}}$. Lower oxygen chemical potential leads to exposed surface metal atoms in the RuO$_2$(110)-Ru surface structure.
 With increasing oxygen chemical potential, the oxygen atoms will cover surface sites, forming the RuO$_2$(110)-O$^{\text{bridge}}$ and RuO$_2$(110)-O$^{\text{cus}}$ structures.

In the bulk structure of RuO$_2$, O is bonded to three Ru atoms, producing the $sp^{\text{2}}$ hybridization, and Ru is bonded to six O atoms, forming $d^{\text{2}}sp^{\text{3}}$ hybridization. From this information, one can predict the catalytically active sites on each surface, since the reactivity of a metal oxide surface is related to the undercoordinated metal and O atoms \cite{cox96}. Observing the RuO$_2$(110)-Ru surface, the five-fold, coordinatively unsaturated site Ru$^{\text{cus,5f}}$ and the four-fold bridge Ru$^{\text{bridge,4f}}$ atoms are undercoordinated. 
This makes the surface Ru atoms catalytically active. Also, even though O$^{\text{3f}}$ is fully bonded with three Ru atoms, other works show that it could be a site for weak hydrogen bonding \cite{over12p3356}. On the RuO$_2$(110)-O$^{\text{bridge}}$ surface, one can observe that while Ru$^{\text{bridge,6f}}$ is fully coordinated, the other surface atoms are not, keeping them catalytically active. For the RuO$_2$(110)-O$^{\text{cus}}$ surface, all the Ru atoms are fully coordinated, but O$^{\text{bridge}}$ and O$^{\text{cus}}$ atoms are undercoordinated, 
bonding with only two and one Ru atoms respectively.

Before studying surface adsorption, the bulk RuO$_2$ is relaxed using the variable cell relaxation method and the GGA functional. The cell parameters of bulk RuO$_2$ are calculated to be $a$ = $b$ = 4.47~\AA, $c$ = 3.08~\AA\, which agree well with experimental X-ray diffraction values of $a$ = $b$ = 4.49~\AA, $c$ = 3.10~\AA\ \cite{atanasoska88p142}.
Using bulk in-plane lattice parameters, a symmetric surface slab structure is constructed using three layers of Ru.
 For the surface, the relaxed Ru$^{\text{bridge,6f}}$-O$^{\text{bridge}}$ length is 1.88~\AA\ and Ru$^{\text{cus,6f}}$-O$^{\text{cus}}$ is 1.68~\AA.
The $z$-axis cell parameter for the slab is set to 30~\AA\, which separates the benzene molecules on the top and  bottom surfaces by at least 14~\AA\ for all cases. This distance ensures that there is no interaction between benzene molecules above and below the slab.

The adsorption energies are calculated for every adsorption relaxation:
\begin{equation}
E_{\text{ads}}=E_{\text{slab$+$molecule}}-E_{\text{slab}}-E_{\text{molecule}},
\label{equation1}
\end{equation}
where $E_{\text{ads}}$ is the adsorption energy, $E_{\text{slab+molecule}}$ is the total energy of the relaxed structure of the slab with the benzene molecule, $E_{\text{slab}}$ is the energy of the relaxed slab structure, and $E_{\text{molecule}}$ is the energy of molecule overlayer with same periodicity as the adsorbed system. The negative value of the adsorption energy indicates that adsorption is energetically favorable.

\subsection{Surface adsorption sites}
\begin{figure}[ht]
\caption{Site naming scheme for all three RuO$_2$(110) surfaces. The figures are all shown in top view. (a) Site naming for RuO$_2$(110)-Ru surface (b) RuO$_2$(110)-O$^{\text{bridge}}$ (c) RuO$_2$(110)-O$^{\text{cus}}$ (d) figure with benzene placed on each site on the surface of RuO$_2$(110)-Ru. Each blue square with arrow indicates a surface oxygen that is added, going from reduced to oxidizied forms of the surface.}
\begin{center}
\includegraphics[scale=0.3,angle=0]{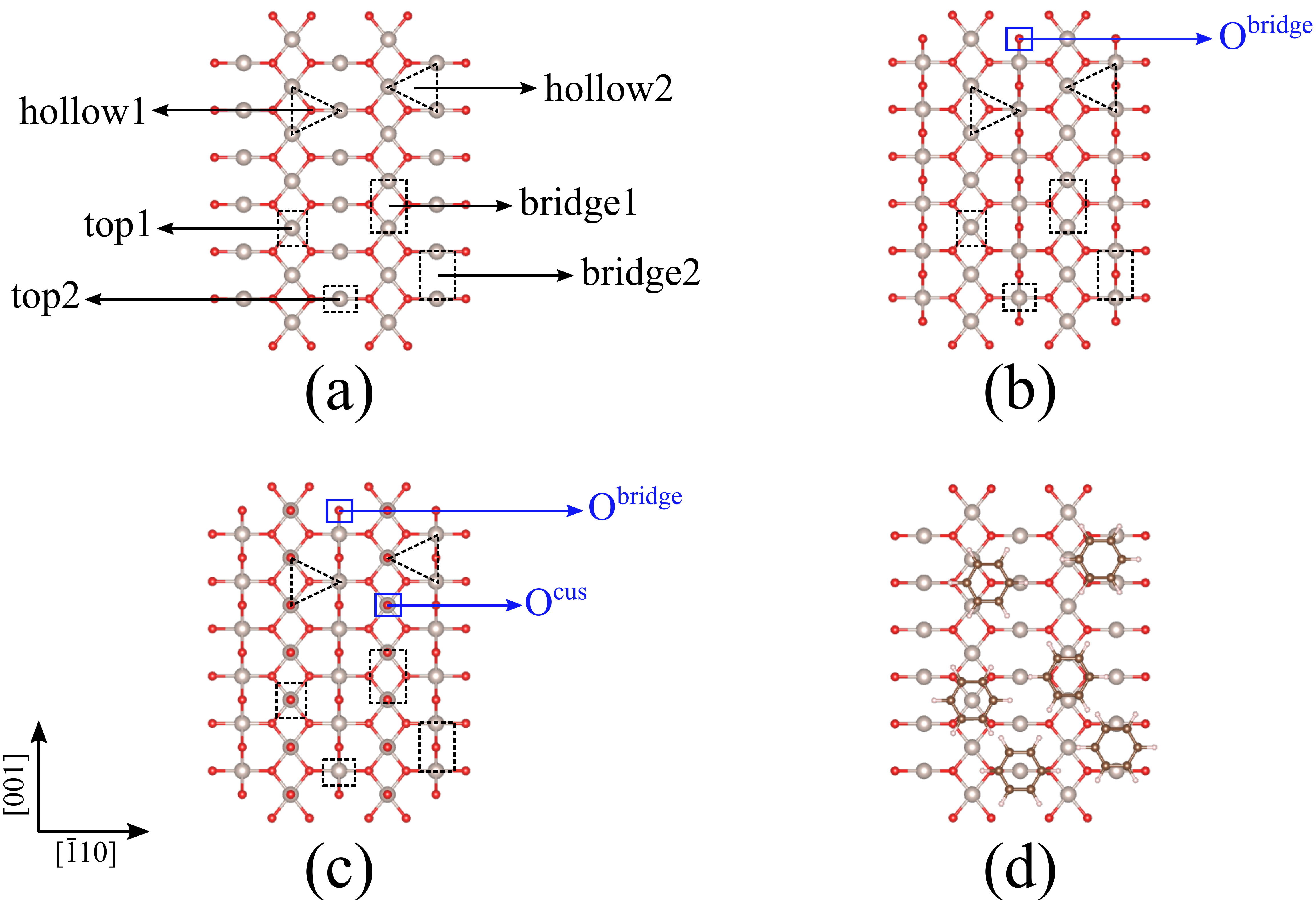}
\end{center}
\label{figure2}
\end{figure}

Three types of adsorption sites, hollow, bridge and top, are determined based on the benzene position relative to the surface Ru atoms. The hollow site is where the three surface Ru atoms are covered by the benzene C atoms. The bridge site is where the benzene C-C bond is on top of the Ru atom. And the top sites are where the center of the benzene ring is directly above the Ru atom. We label six distinct sites for each surface as hollow1, hollow2, top1, top2, bridge1, bridge2. 
The benzene molecules are initially placed 3~\AA\ above the Ru atoms of the RuO$_2$(110) surface.
 The lateral starting coordinates (Fig. \ref{figure2}) are labelled as follows.
 
The hollow sites are located with the center of the benzene ring directly on top of a Ru three-fold hollow site. Specifically, the hollow1 site has a O$^{\text{3f}}$ atom inside the carbon ring, whereas hollow2 is without an O$^{\text{3f}}$ atom in the carbon ring. The top sites are placed so that the center of the benzene ring is directly on top of a particular surface atom, depending on the surface structure. For the RuO$_2$(110)-Ru and RuO$_2$(110)-O$^{\text{bridge}}$ surfaces, the top1 site has the benzene ring centered over the Ru$^{\text{cus,5f}}$ atom, and for the RuO$_2$(110)-O$^{\text{cus}}$ surface the top1 site has the benzene ring over the O$^{\text{cus}}$ atom, which is directly above the Ru atom, (this O$^{\text{cus}}$ is marked as blue in Fig. 2c). For the RuO$_2$(110)-Ru surface, the top2 site is over the 
Ru$^{\text{bridge,4f}}$ atom, and for the RuO$_2$(110)-O$^{\text{bridge}}$ and RuO$_2$(110)-O$^{\text{cus}}$ surfaces, the top2 is over the Ru$^{\text{bridge,6f}}$ atom. The bridge sites have the center of the benzene ring directly above the midpoint of two surface Ru atoms. Bridge1 is between adjacent Ru$^{\text{cus}}$ atoms, and bridge2 is between Ru$^{\text{bridge}}$ atoms. For the RuO$_2$(110)-O$^{\text{bridge}}$ and RuO$_2$(110)-O$^{\text{cus}}$ surfaces, the bridge2 site contains O$^{\text{bridge}}$ connecting the Ru$^{\text{bridge}}$ atoms, marked as blue in Fig. 2b.
The schematic of site naming is shown in Fig. \ref{figure2}, and benzene molecules are placed so that two of the C-C bonds would be along [$\bar{1}$10].

To choose the periodic unit cell size for our calculations, we calculate the nearest distance between two hydrogen atoms on nearby benzene molecules in the periodic 
$1\times1$, $1\times2$, and $\sqrt{5}\times\sqrt{5}$ cells. For the case of one benzene per $1\times1$ cell, the nearest distance is less than 1~\AA,\ which is unreasonable. The $1\times2$ and $\sqrt{5}\times\sqrt{5}$ cells give minimum distances of 1.34~\AA\ and 4.11~\AA\ between two hydrogen atoms on adjacent benzenes, respectively. Accordingly, $1\times2$ and $\sqrt{5}\times\sqrt{5}$ surface cells are modeled in this study. Additionally, to compare and confirm the results of $\sqrt{5}\times\sqrt{5}$ cells, we also include the $2\times3$ surface adsorption calculations to represent low coverage.

\section{Results}

\begin{table}[h]
\begin{center}
\begin{tabular}{l ccc}
\hline
\hline
 & RuO$_2$(110)-Ru/bz (eV)    &  RuO$_2$(110)-O$^{\text{bridge}}$/bz (eV)  &  RuO$_2$(110)-O$^{\text{cus}}$/bz (eV)\\
\hline
hollow1  &   -0.20   & -0.23  & -0.06\\
hollow2 &    -0.39   & -0.23  & -0.08\\
top1       &   -0.20   & -0.24  & -0.06 \\
top2       &   -0.05   & -0.01  & -0.08\\
bridge1  &   -0.14   & -0.22  & -0.05\\
bridge2  &   -0.03   & -0.01  & -0.09\\
\hline
\hline
\label{table1}
\end{tabular}
\caption{Adsorption energies of benzene(bz) on a periodic $1\times2$ surface of RuO$_2$(110)-Ru, RuO$_2$(110)-O$^{\text{bridge}}$, and RuO$_2$(110)-O$^{\text{cus}}$ using the GGA functional.}
\end{center}
\end{table}

\begin{table}[h]
\begin{center}
\begin{tabular}{l ccc}
\hline
\hline
 & RuO$_2$(110)-Ru/    &  RuO$_2$(110)-O$^{\text{bridge}}$/  &  RuO$_2$(110)-O$^{\text{cus}}$/\\
 & hollow2 (eV)  &  top1 (eV)   &  bridge2 (eV)\\
\hline
$1\times2$-GGA & -0.39  &  -0.24  & -0.09\\
$1\times2$-TS &  -1.40  &  -0.66  &  -0.65\\
$\sqrt{5}\times\sqrt{5}$-GGA & -0.63  & -0.28  & -0.13\\
$\sqrt{5}\times\sqrt{5}$-TS & -1.68  & -1.06  & -0.82\\
$2\times3$-GGA & -0.63  &  -0.30  & -0.15\\
$2\times3$-TS &  -1.61  &  -1.06  & -0.84\\
\hline
\hline
\label{table2}
\end{tabular}
\caption{Adsorption energies of benzene with $1\times2$, $\sqrt{5}\times\sqrt{5}$, and $2\times3$ periodicities on the RuO$_2$(110)-Ru hollow2, RuO$_2$(110)-O$^{\text{bridge}}$ top1, and RuO$_2$(110)-O$^{\text{cus}}$ bridge2 using the GGA functionals and vdW corrected values with TS method. Note that the large difference between GGA and vdW corrected energies demonstrates that the long-range vdW attraction plays a major role in benezene adsorbs on RuO$_2$ surface.}
\end{center}
\end{table}

\subsection{Energy and geometry of benzene adsorption on RuO$_2$(110) surface}

\begin{figure}[ht]
\caption{Projected density of states of benzene adsorbed on the $1\times2$ supercell of the (a) RuO$_2$(110)-Ru and (b) RuO$_2$(110)-O$^{\text{cus}}$ surfaces. The C atom orbitals on the benzene ring and the closest Ru surface atom orbitals are shown. Note the orbital interaction between the $p_z$ orbitals of C and $d_y$$_z$ orbital of Ru near the Fermi energy. The C atom that is closest to the surface and its closest surface O atom orbitals are shown. Note the C atom orbital of benzene is close to an isolated molecule, indicating that no chemical bond forms, and hence presenting physisorptive phenomenon on the oxidized surface.}
\begin{center}
\includegraphics[scale=0.17,angle=0]{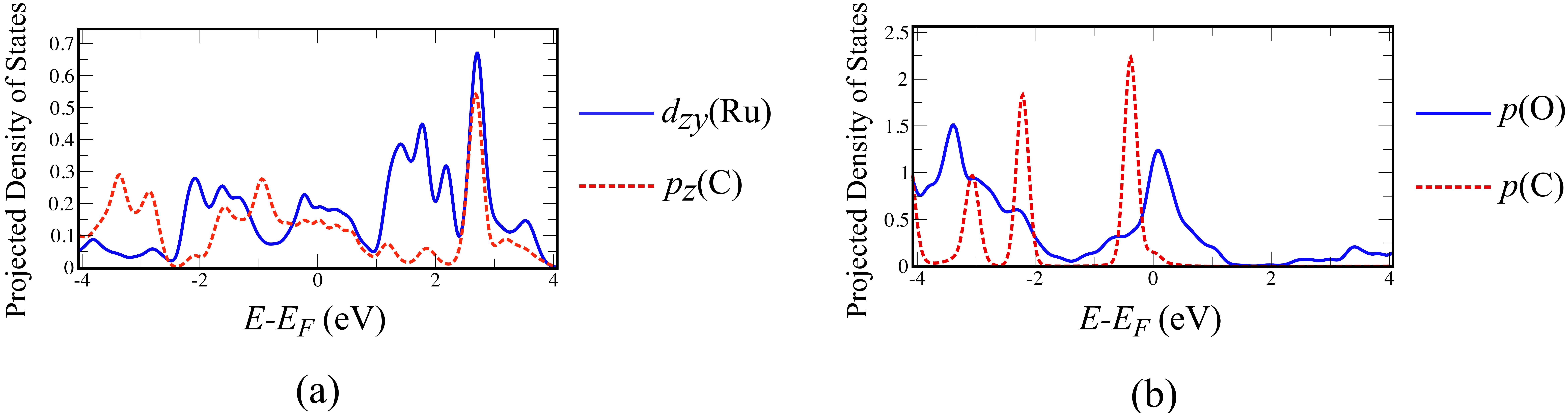}
\end{center}
\label{figure3}
\end{figure}

We start by investigating the adsorption of benzene on the RuO$_2$(110) surface using the GGA functional. 
The relaxation pathways of benzene on the most reduced surface, RuO$_2$(110)-Ru, using a $1\times2$ periodic cell are quite interesting and highlight important adsorption interactions. The site with the strongest adsorption energy is the hollow2 site, since in this position the C atoms of benzene and Ru surface atoms can maximize their strong attraction without much interference from surface O atoms. The strong interactions of C atom $p$$_z$ orbitals and the Ru $d$$_y$$_z$ orbitals are shown in Fig. \ref{figure3}.

 The adsorption relaxation of benzene on the hollow2 site of RuO$_2$(110)-Ru surface follows three steps. The first step is the attraction of C and Ru. This makes the benzene molecule tilt so that two C atoms get closer to the two Ru$^{\text{bridge,4f}}$ atoms. The benzene H atoms do not show much attraction toward the surface at this point. The second observed behavior is the slight rotation of benzene, by $\approx$ 11.9$^{\circ}$.
 The third step strengthens adsorption, as the benzene molecule moves closer to the surface. In this step, in addition to the two C already attracted to the Ru$^{\text{bridge,4f}}$ atoms, the C on the other side of the ring is attracted toward the Ru$^{\text{cus,5f}}$. Also, the H atoms of the benzene ring are slightly attracted towards the O$^{\text{3f}}$ atoms. At this step, the benzene molecule deforms significantly.

Other than adsorption on hollow2 of the RuO$_2$(110)-Ru surface in a $1\times2$ cell, all the other sites show a similar behavior. For the other sites, the first step is the rotation of benzene by $\approx$ 12$^{\circ}$. The second step is the benzene tilting by $\approx$ 20$^{\circ}$.
 For all cases on the RuO$_2$(110)-Ru surface, the attraction of C and surface Ru atoms (Ru$^{\text{cus,5f}}$ and Ru$^{\text{bridge,4f}}$) is observed, but it is not as strong as hollow2. Also, some H atoms on the benzene are attracted to the O$^{\text{3f}}$ atoms on the surface, as was predicted from previous DFT work on O$^{\text{3f}}$ as a hydrogen bonding site\cite{over12p3356}. On all sites of RuO$_2$(110)-Ru including the hollow2, during and after the adsorption, the molecule does not move from its initial site to find its most favorable site. Instead it generally stays on its initial site, and the largest lateral movement that occurs is the rotation of the molecule. 

To study the chemisorption and physisorption mechanisms with less intermolecular interaction, the adsorption is studied in a $\sqrt{5}\times\sqrt{5}$ supercell of the RuO$_2$(110)-Ru slab using the GGA functional. For the hollow2 site, the final structure shows a very similar geometry to the $1\times2$, but with significantly enhanced chemisorption energy of -0.63 eV. Also, there is no rotation of benzene during the relaxation. For the top1 site, one of the most strongly physisorbed (-0.20 eV) sites on the $1\times2$ cell of RuO$_2$(110)-Ru surface, the calculated adsorption energy on the $\sqrt{5}\times\sqrt{5}$ supercell increases (-0.27 eV). This stronger adsorption energy with lower coverage indicates a reduction of molecular interaction and increased interaction with the surface. Since the changes of the adsorption energies from a $1\times2$ to a $\sqrt{5}\times\sqrt{5}$ cell are significant, the adsorption of benzene in a $2\times3$ cell is also calculated. The geometry and the energy in the $2\times3$ supercell are very similar to the $\sqrt{5}\times\sqrt{5}$ surface in all cases.

\begin{figure}[ht]
\caption{Top and side views of the equilibrium structure of benzene adsorbed on the hollow2 site of the RuO$_2$(110)-Ru surface with different periodic cell sizes using GGA  functionals. (a) $1\times2$ supercell of the RuO$_2$(110)-Ru surface using the GGA functional. (b) $\sqrt{5}\times\sqrt{5}$ supercell of the RuO$_2$(110)-Ru surface using the GGA functional. vdW calculations using DFTD2 and TS have similar geometries.}
\begin{center}
\includegraphics[scale=0.18,angle=0]{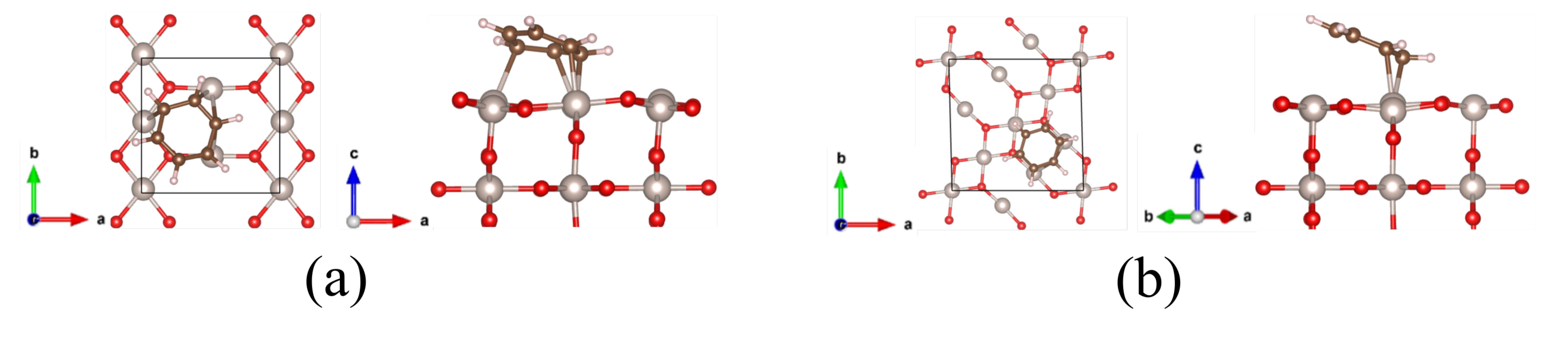}
\end{center}
\label{figure4}
\end{figure}

For further investigation of chemisorption on the reduced surface, we account for the van der Waals (vdW) interaction in the calculation by using DFTD2~\cite{Grimme06p1787} and TS~\cite{Tkatchenko09p073005} methods. 
In the periodic $1\times2$ and $\sqrt{5}\times\sqrt{5}$ supercell, the benzene adopts similar geometries regardless of the inclusion of vdW correction, as shown in Fig. \ref{figure4}. The hollow2 site proves to be a chemisorption site because of the strong carbon-metal bonds. For confirmation, benzene adsorption is also examined in the $2\times3$ supercell using the DFTD2 and TS methods. The geometric and energetic results are very similar to those on the $\sqrt{5}\times\sqrt{5}$ supercell.


On the oxidized surfaces (RuO$_2$(110)-O$^{\text{bridge}}$ and RuO$_2$(110)-O$^{\text{cus}}$), O atoms gain electrons and Ru atoms lose electrons. This makes the O atoms interfere with the Ru-C attraction, weakening it. In addition, the O$^{\text{bridge}}$ and O$^{\text{cus}}$ atoms attract the H atoms of the benzene, forming weak hydrogen bonds, as also shown in other DFT studies\cite{sun04p235402}.
We again start our investigation using the GGA functional on the $1\times2$ cell. The relaxation pathways for benzene on the $1\times2$ cell of oxidized surfaces are very similar regardless of the site, suggesting physisorption. 
 The first step is the repulsion of benzene from the surface without any changes to its geometry, rotation, or tilt. The extent of the benzene displacement away from the surface varies among sites from 0.5~\AA\ to 2~\AA. This repulsion arises from the interaction between the surface O atoms and C atoms on the benzene, also shown by the projected density of states in FIG \ref{figure3}.
 The second step consists of benzene molecule rotation by $\approx$ 25$^{\circ}$, induced by intermolecular interactions. The last step is the molecular tilting, which maximizes the H bonding between the surface O atoms and H atoms of benzene. The tilting is $\approx$ 46$^{\circ}$, which is greater than on the RuO$_2$(110)-Ru surface. Again, the benzene molecules do not move around laterally, but stay on their initial site.
Besides the relaxation pathways, a minor difference between the two oxidized surfaces is that the RuO$_2$(110)-O$^{\text{cus}}$ surface has a slightly weaker benzene binding energy than the RuO$_2$(110)-O$^{\text{bridge}}$.

\begin{figure}[ht]
\caption{Top and side views of the final structure of benzene adsorbed on the bridge2 site of the RuO$_2$(110)-O$^{\text{cus}}$ surface of different periodic cell sizes using GGA. (a) $1\times2$ supercell (b) $\sqrt{5}\times\sqrt{5}$ supercell. vdW calculations using DFTD2 and TS show the similar geometries.}

\begin{center}
\includegraphics[scale=0.18,angle=0]{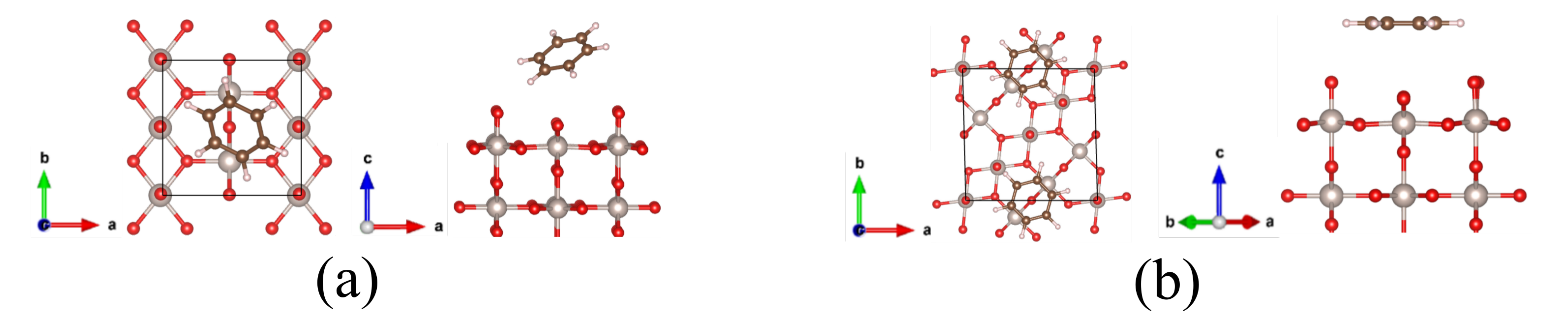}
\end{center}
\label{figure5}
\end{figure}

To further study the adsorption on oxidized RuO$_2$ with reduced intermolecular interaction, the sites with the most stable energies in the $1\times2$ cell were chosen for study in $\sqrt{5}\times\sqrt{5}$ cells (top1 of RuO$_2$(110)-O$^{\text{bridge}}$ and bridge2 of RuO$_2$(110)-O$^{\text{cus}}$), using the GGA functional. The GGA adsorption energy in the $\sqrt{5}\times\sqrt{5}$ cell of the RuO$_2$(110)-O$^{\text{bridge}}$ top1 surface is -0.28 eV, and for RuO$_2$(110)-O$^{\text{cus}}$ bridge2 it is -0.13 eV. During the relaxation on this larger cell, intermolecular interaction is weak, so the molecules do not tilt or rotate.
In addition, to observe the effects of vdW interactions on oxidized RuO$_2$, the DFTD2 and TS methods are applied to the $1\times2$, $\sqrt{5}\times\sqrt{5}$, and $2\times3$ cells of the top1 site of the RuO$_2$(110)-O$^{\text{bridge}}$ surface and the bridge2 site of the RuO$_2$(110)-O$^{\text{cus}}$ surface. The trends of both the energy and the geometry of the relaxed structure conform to the GGA results for oxidized RuO$_2$(110) surfaces, as shown in Fig. \ref{figure5}.

\subsection{Adsorption structure stability phase diagram}
\begin{figure}[ht]
\caption{The benzene adsorption surface phase diagram of the RuO$_2$(110) surface. Each color indicates a distinct surface adsorption condition. The legend shows both O and benzene coverage. The partial pressure scales for benzene and oxygen are at 300~K. }
\begin{center}
\includegraphics[scale=0.93,angle=0]{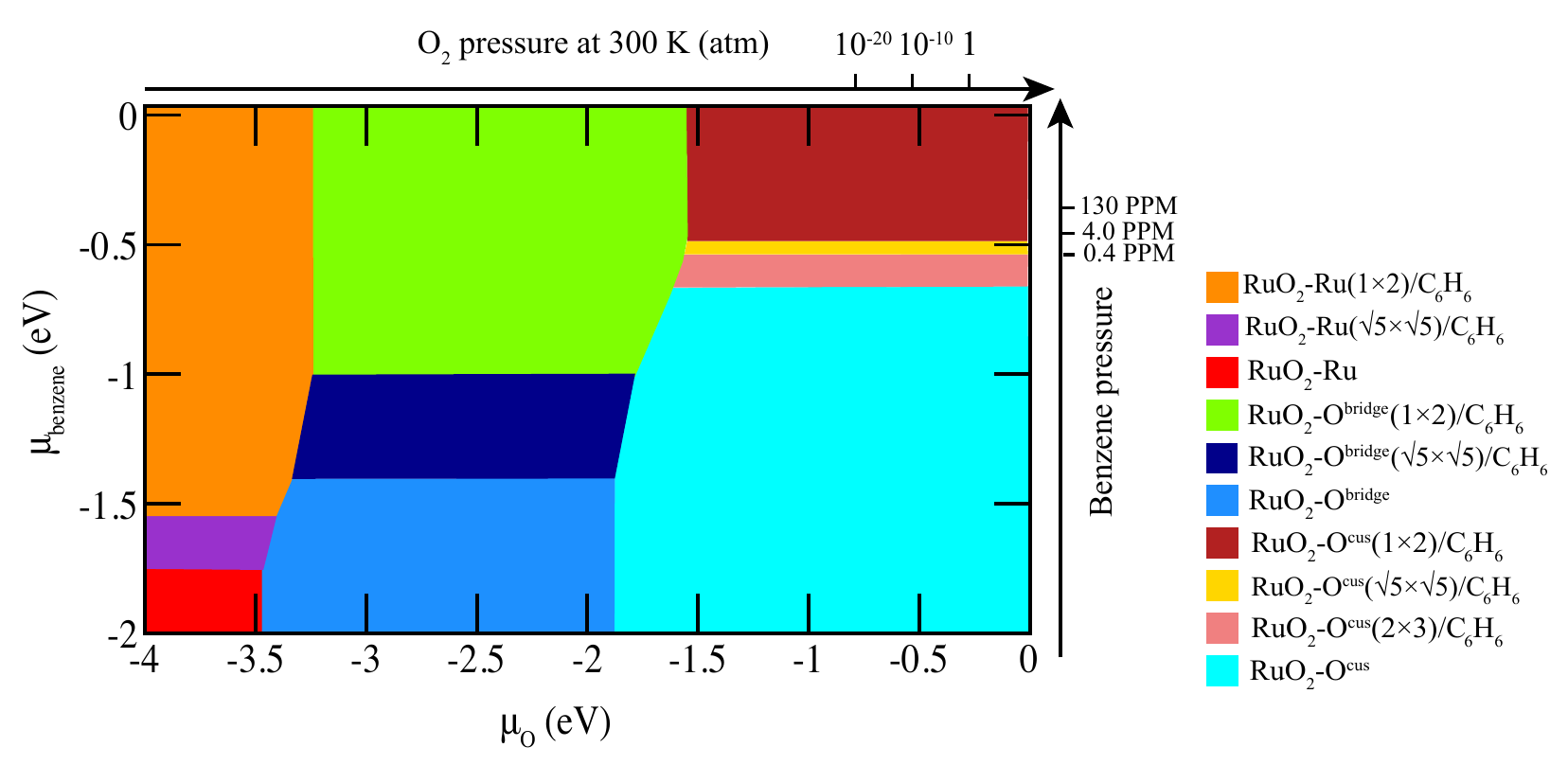}
\end{center}
\label{figure6}
\end{figure}

The preferred coverages that minimize the Gibbs free energy of adsorption for varying oxygen and benzene chemical potentials are given in Fig. \ref{figure6}. 
We express the Gibbs free energy in terms of benzene and oxygen chemical potentials as following~\cite{Rogal07p205433}, 
\begin{equation}
\Delta G^{\text{ads}}(\Delta\mu_{\text{O}},\Delta\mu_{\text{bz}})=-\frac{1}{A}\Delta E_{\text{O,bz@slab}}^{\text{ads}}+\frac{N_{\text{O}}}{A}\Delta\mu_{\text{O}}+\frac{N_{\text{bz}}}{A}\Delta\mu_{\text{bz}},
\end{equation}
where $\Delta G^{\text{ads}}(\Delta\mu_{\text{O}},\Delta\mu_{\text{bz}})$ is the benzene adsorption Gibbs free energy as a function of chemical potential of oxygen ($\Delta\mu_{\text{O}}$) and benzene ($\Delta\mu_{\text{bz}}$), $A$ is the surface area of the slab, $\Delta E_{\text{O,bz@slab}}^{\text{ads}}$ is the adsorption energy of O atoms and benzene molecules on the reduced surface, $N_{\text{O}}$ is the number of O atoms, and $N_{\text{bz}}$ is the number of benzene molecules. The chemical potential of benzene and oxygen corresponds to pressure which can be expressed as, 
\begin{equation}
\Delta\mu_{\text{bz}}=\mu_{\text{bz}}^{\text{0}}+k_BT\ln\Big({\frac{p_{\text{bz}}}{p_0}}\Big),
\end{equation}
and
\begin{equation}
\Delta\mu_{\text{O}}=\mu_{\text{O}}^{\text{0}}+\frac{1}{2}k_BT\ln\Big({\frac{p_{\text{O}_2}}{p_0}}\Big),
\end{equation}
where $\mu_{\text{bz}}^{\text{0}}$ and $\mu_{\text{O}_2}^{\text{0}}$ are the standard chemical potential of benzene and oxygen gas molecules, and $p_0$ is the pressure of 1 atm. The standard chemical potential for oxygen can be obtained from the experimental thermodynamic table~\cite{stull1971janaf}. For the standard chemical potential of benzene, the value is approximated from experimental standard Gibbs free energy~\cite{schwarzenbach2005environmental}. 
In the surface phase diagram, we plot the stability ranges of benzene overlayers on RuO$_2$(110) with different periodicities and of extents of oxidation. The pressure corresponding to the chemical potential is shown assuming room temperature ($T$=$300$ K). 
In Fig. \ref{figure6}, we show the surface phase diagram based on DFTD2 calculations. Starting from the lower left, where the chemical potentials of both benzene and oxygen are very low, RuO$_2$(110)-Ru is the most stable surface. As increases to the right, the most stable phase goes from RuO$_2$(110)-Ru, to RuO$_2$(110)-O$^{\text{bridge}}$, and then to RuO$_2$(110)-O$^{\text{cus}}$, which is consistent with previous findings \cite{Reuter01p035406}.
 Along the benzene chemical potential axis at low $\mu_{\text{O}}$, the RuO$_2$(110)-Ru surface with benzene adsorption at different coverages is shown. As $\mu_{\text{benzene}}$ increases, higher coverage is favored.
 Note that 2$\times$3 surface adsorption phase for RuO$_{\text{2}}$-Ru and RuO$_{\text{2}}$-O$^{\text{bridge}}$ does not appear on the graph due to almost equivalent adsorption condition and energetics with $\sqrt5\times\sqrt5$ periodic cell.
 Moving diagonally up and to the right, we can observe the change of the surface termination (RuO$_2$(110)-Ru to RuO$_2$(110)-O$^{\text{cus}}$). At high chemical potential of both species, we can observe increase of the benzene coverage on the RuO$_2$(110)-O$^{\text{cus}}$ surface (RuO$_2$(110)-O$^{\text{cus}}$($2\times3$) to RuO$_2$(110)-O$^{\text{cus}}$($1\times2$)). The low coverage $2\times3$ surface with one benzene per six primitive cells is only favorable on the RuO$_2$(110)-O$^{\text{cus}}$ surface. The top right corner indicates that when both $\mu_{\text{benzene}}$ and $\mu_{\text{O}}$ are high, the most stable surface is RuO$_2$(110)-O$^{\text{cus}}$ with high molecular coverage. The scale for benzene pressure is converted to the experimental values in unit of PPM~\cite{Brand13p1248}. The PPM units of pressure of benzene gas is converted based on the ideal gas law as shown in following~\cite{de2010air},
 \begin{equation}
 1~\text{PPM} = \frac{1 \mu \text{moles gas}}{1 \text{mole air}} = \frac{V_n}{M}\frac{1 \mu \text{g gas}}{1 \text{L air}},
 \end{equation}
where $T$ is temperature, $M$ is the molar mass of the gas, and $V_n$ is the molar volume of the gas taking the form of
\begin{equation}
V_n = \frac{RT}{p}.
\end{equation}
At the pressure, $p=1$ atm, and room temperature (300 K), 1 PPM = 3.151 $\times$ 10$^{-7}$ atm for benzene gas.


 
Based on the surface phase diagram, we may conclude that for most of the experimental conditions, the RuO$_2$(110)-O$^{\text{cus}}$ surface is the stable phase. And with such a fully oxidized surface, the density of the adsorbed benzene layer depends on the benzene pressure.

\section{Conclusion}
We investigate the surface adsorption of benzene on RuO$_2$(110). The adsorption mechanism of benzene on RuO$_2$(110) depends on the site, O coverage, and the benzene coverage. Clear trends emerge: the C and Ru atoms attract strongly, while benzene H and oxide O atoms have a weaker attraction. Also, increasing the benzene coverage makes the molecules interact with each other, reducing surface bonding. As the surface becomes more oxidized, the O surface atoms make the benzene adsorption weaker. Benzene can strongly chemisorb to the Ru-terminated surface. However, full surface oxidation leads to benzene physisorption. With understanding of benzene adsorption geometries, energies, and their relation to the surface oxidation, we can develop realistic models of the formation of tribopolymer on metal oxide contacts and provide useful information for further dynamical studies.

\section{Acknowledgments}
H.D.K. was supported by the University of Pennsylvania NSF/Louis Stokes Alliance for Minority Participation (LSAMP) program.
J.Y. was supported by the U.S. National Science Foundation, under grant CMMI-1334241.
Y.Q. was supported by the U.S. National Science Foundation, under grant DMR-1124696.
A.M.R. was supported by the U.S. Department of Energy, under grant DE-FG02-07ER15920.
Computational support is provided by the HPCMO of the U.S. DOD and the NERSC of the U.S. DOE.

\bibliography{rappecites}

\end{document}


The adsorption energies with different van der Waals (vdW) corrections are listed in the following tables. Tab \ref{Tab:1} collected the energies obtained from geometric optimizations with different vdW corrections, whereas Tab \ref{Tab:2} shows the energies based on GGA optimized structure but only correcting the energy. In general, DFTD3~\cite{Grimme10p154104} and TS~\cite{Tkatchenko09p073005} corrected energies are in close match with each other. Also, TS correction is proved to better predict the surface adsorption system geometrically and energetically~\cite{Ruiz12p146103}. Energies obatined from DFTD2 geometric optimized calculation is generally lower compare to that from TS, because DFTD2 correction with empirical long-range coefficients tends to overestimate the binding between the adsorbate and the surface~\cite{Chen12p424211}.   
\begin{table}[h]
\begin{center}
\begin{tabular}{l ccc}
\hline
\hline
 & RuO$_2$(110)-Ru/    &  RuO$_2$(110)-O$^{\text{bridge}}$/  &  RuO$_2$(110)-O$^{\text{cus}}$/\\
 & hollow2 (eV)  &  top1 (eV)   &  bridge2 (eV)\\
\hline
$1\times2$-GGA & -0.39  &  -0.24  & -0.09\\
$1\times2$-TS &  -1.40  &  -0.66  &  -0.65\\
$1\times2$-DFTD2 &  -1.58  &  -1.18  & -0.58 \\
$\sqrt{5}\times\sqrt{5}$-GGA & -0.63  & -0.28  & -0.13\\
$\sqrt{5}\times\sqrt{5}$-TS & -1.68  & -1.06  & -0.82\\
$\sqrt{5}\times\sqrt{5}$-DFTD2 & -1.73  & -1.42  & -0.62\\
$2\times3$-GGA & -0.63  &  -0.30  & -0.15\\
$2\times3$-TS &  -1.61  &  -1.06  & -0.84\\
$2\times3$-DFTD2 &  -1.71  &  -1.44  & -0.65\\
\hline
\hline
\end{tabular}
\caption{Adsorption energies of benzene with $1\times2$, $\sqrt{5}\times\sqrt{5}$, and $2\times3$ periodicities on the RuO$_2$(110)-Ru hollow2, RuO$_2$(110)-O$^{\text{bridge}}$ top1, and RuO$_2$(110)-O$^{\text{cus}}$ bridge2 using the GGA functional and energies with TS and DFTD2 vdW corrections. Note that the vdW corrections is added during the geometric optimization.}
\label{Tab:1}
\end{center}
\end{table}
 
\begin{table}[h]
\centering
\caption{Adsorption energies of benzene with different molecular surface coverage using DFTD2, DFTD3 and TS vdW corrections. the values from all three methods are very similar. The D3 and TS correction on adsorption energy shows a little better agreement compare to that with D2 corrections. Note the GGA optimized structures are used to obtain the energies.}
\label{Tab:2}
\begin{tabular}{l ccc}
\hline
\hline
 & RuO$_2$(110)-Ru/    &  RuO$_2$(110)-O$^{\text{bridge}}$/  &  RuO$_2$(110)-O$^{\text{cus}}$/\\
 & hollow2 (eV)  &  top1 (eV)   &  bridge2 (eV)\\
 \hline
$1\times2$-GGA   & -0.39                      & -0.24                     & -0.09                        \\
$1\times2$-DFTD2    & -1.64                      & -0.94                     & -0.40                        \\
$1\times2$-DFTD3    & -1.28                      & -0.62                     & -0.42                        \\
$1\times2$-TS    & -1.30                      & -0.67                     & -0.37                        \\
$\sqrt{5}\times\sqrt{5}$-GGA & -0.63                      & -0.28                     & -0.13                        \\
$\sqrt{5}\times\sqrt{5}$-DFTD2  & -1.67                      & -0.84                     & -0.50                        \\
$\sqrt{5}\times\sqrt{5}$-DFTD3  & -1.69                      & -0.82                     & -0.53                        \\
$\sqrt{5}\times\sqrt{5}$-TS  & -1.67                      & -0.84                     & -0.55                        \\
$2\times3$-GGA   & -0.63                      & -0.30                     & -0.15                        \\
$2\times3$-DFTD2    & 1.59                       & -0.82                     & -0.73                        \\
$2\times3$-DFTD3    & -1.60                      & -0.83                     & -0.75                        \\
$2\times3$-TS    & -1.60                      & -0.85                     & -0.71 \\
\hline
\hline
\end{tabular}
\end{table}
\clearpage
\bibliography{rappecites}